\documentclass[conference]{IEEEtran}
\IEEEoverridecommandlockouts
\usepackage{cite}
\usepackage{amsmath,amssymb,amsfonts}
\usepackage{algorithmic}
\usepackage{graphicx}
\usepackage{textcomp}
\usepackage{xcolor}
\usepackage{url}
\usepackage{tabularx}
\def\BibTeX{{\rm B\kern-.05em{\sc i\kern-.025em b}\kern-.08em
    T\kern-.1667em\lower.7ex\hbox{E}\kern-.125emX}}
\begin{document}

\title{An AI-Assisted Migration Framework for Transforming Legacy Scientific Applications into Reusable Cloud-Based Workflows\\
}



\author{
\IEEEauthorblockN{
Nafiseh Soveizi\IEEEauthorrefmark{1},
Sven Tesselaar\IEEEauthorrefmark{1},
Hero Robinson Brouwer\IEEEauthorrefmark{1},
Zhiming Zhao\IEEEauthorrefmark{1}\IEEEauthorrefmark{2}
}

\IEEEauthorblockA{
\IEEEauthorrefmark{1}
University of Amsterdam, Science Park 904, Amsterdam, the Netherlands\\
\{n.soveizi, z.zhao\}@uva.nl
}

\IEEEauthorblockA{
\IEEEauthorrefmark{2}
LifeWatch ERIC Virtual Lab Innovation Center, Science Park 904, Amsterdam, the Netherlands
}
}



\maketitle

\begin{abstract}

Legacy scientific applications remain valuable research assets but are often tightly coupled to project-specific execution environments, limiting their reuse, reproducibility, and deployment within modern scientific workflow systems and cloud-native Virtual Research Environments (VREs). Existing migration approaches primarily target individual artifacts, such as notebooks or containers, and provide limited support for systematically transforming heterogeneous legacy applications into reusable cloud-native workflows. This paper presents an AI-assisted migration framework that combines the Reference Model of Open Distributed Processing (RM-ODP)-guided architectural analysis, Large Language Models (LLMs), and Design Structure Matrix (DSM) analysis. The framework first uses RM-ODP to guide an LLM in identifying reusable workflow components, their interfaces, and execution dependencies from heterogeneous legacy applications. The resulting workflow structure is then iteratively evaluated and refined using DSM analysis. Finally, an LLM-based workflow generator implements the validated workflow components and produces containerized execution environments and executable workflow definitions for deployment in cloud-native workflow systems, including VREs. The framework was evaluated on two legacy scientific applications from different scientific domains. In both cases, the applications were successfully transformed into reusable cloud-native workflows while preserving their original functionality, demonstrating the feasibility of the proposed approach for modernizing legacy scientific software.

\end{abstract}


\section{Introduction}

Scientific workflows have become the primary mechanism for conducting
computational research by orchestrating software components, datasets,
and computational resources into reproducible and scalable analysis
pipelines \cite{deelman2018future,deelman2015pegasus}. Workflow
Management Systems (WMSs) enable researchers to compose, execute, and
share workflows across heterogeneous computing infrastructures, while
cloud-native Virtual Research Environments (VREs) further integrate
workflow execution with shared data, software, and computational
resources to support collaborative and reproducible scientific research
\cite{deelman2015pegasus,goecks2010galaxy,wilkinson2016fair,goble2010myexperiment}.

Despite these advances, much scientific knowledge remains embedded in
legacy research software, including standalone scripts, Jupyter
notebooks, simulation codes, and domain-specific applications
\cite{hannay2009software,wilson2014best,carver2022research}. These
applications were typically developed for project-specific execution
environments rather than reusable workflows. As a result, they often
contain tightly coupled implementations, implicit execution logic,
hard-coded configurations, and undeclared software dependencies,
making them difficult to reuse, reproduce, and migrate to modern
cloud-based workflow platforms.

Migrating legacy software into reusable workflows is a challenging
software engineering task. It requires identifying meaningful workflow
components, reconstructing execution and data dependencies,
externalizing software environments, and preserving the scientific
behaviour of the original application. These activities are largely
performed manually, demanding substantial software engineering and
domain expertise that limits the scalability of migration across
heterogeneous scientific applications.

Existing approaches mainly address isolated migration tasks, such as
notebook containerization, workflow execution, dependency management,
and cloud deployment
\cite{binder,reana,papermill,zhao2022naavre}. While valuable, they
provide limited support for systematically understanding legacy
applications, defining reusable workflow boundaries, and automatically
generating executable workflows. Recent advances in Large Language
Models (LLMs) have demonstrated promising capabilities for software
understanding and code generation, yet few approaches combine LLMs with
established software architecture principles and objective quality
assessment to guide workflow migration.

To address this gap, this paper presents an AI-assisted migration
framework for transforming legacy scientific applications into reusable
cloud-native workflows. The framework first combines the Reference Model of Open Distributed Processing (RM-ODP)~\cite{ISO10746} with a LLM to identify workflow components, their interfaces, and execution dependencies from heterogeneous legacy artifacts. RM-ODP provides complementary architectural viewpoints that capture the functional, informational, computational, engineering, and technological aspects of legacy applications, enabling the LLM to identify reusable workflow components based on architectural responsibilities rather than source-code structure alone. The resulting workflow structure is then evaluated and iteratively refined using Design Structure Matrix (DSM) analysis \cite{eppinger2012}, which provides an objective assessment of component dependencies, coupling, and modularity to identify structural issues that guide workflow refinement. Finally, an LLM-based workflow generator implements the validated
workflow components and produces containerized workflow artifacts and
executable workflow definitions that can be executed in cloud-native
workflow platforms, including VREs.

The proposed framework is evaluated using two representative legacy
applications from different scientific domains: BGIS, an urban digital
twin application implemented as standalone Python scripts, and FreVA, a
food-resource optimization model implemented as a Jupyter notebook.
Together, these case studies demonstrate that the proposed approach can
systematically migrate heterogeneous legacy software into reusable
cloud-native workflows while preserving the functionality of the
original applications.





The remainder of this paper is organized as follows. Section~II reviews
scientific workflow systems, VREs, and existing
approaches to legacy workflow migration. Section~III presents the
research methodology and the proposed framework. Section~IV reports the
experimental evaluation. Section~V discusses the findings, lessons
learned, and limitations. Finally, Section~VI concludes the paper and
outlines future research directions.

\section{Background and Related Work}

\subsection{Scientific Workflow Systems and Virtual Research Environments}

Scientific WMSs provide computational environments for composing, executing, and monitoring scientific analyses across distributed computing infrastructures. By explicitly representing computational tasks and their dependencies, they improve reproducibility, portability, and scalability of scientific applications \cite{deelman2015pegasus,deelman2018future}. Mature workflow systems, including Pegasus~\cite{deelman2015pegasus}, 
Galaxy~\cite{goecks2010galaxy}, 
Kepler~\cite{ludascher2006scientific,altintas2004kepler}, 
Nextflow~\cite{di2017nextflow}, and Snakemake~\cite{koster2012snakemake}, 
support workflow orchestration, provenance management, and execution on 
cloud and high-performance computing infrastructures.

VREs extend workflow systems by integrating workflow execution with shared datasets, computational resources, metadata management, and collaborative services within a unified research platform. Rather than treating software as isolated scripts, VREs enable researchers to share reusable computational workflows and scientific assets across research communities.

Notebook-based scientific computing has become increasingly popular because Jupyter notebooks combine executable code, documentation, and visualisation within a single environment \cite{kluyver2016}. However, notebooks were originally designed for interactive development rather than scalable and reusable workflow execution. Notebook-as-a-VRE (NaaVRE) addresses this limitation by extending Jupyter notebooks with automatic feature extraction, component containerisation, workflow composition, and cloud-native execution \cite{zhao2022naavre}. This enables notebook cells to be transformed into reusable workflow components while preserving the interactive development experience familiar to researchers.

\subsection{Legacy Scientific Workflow Migration}

Although workflow platforms provide mature execution environments, a large proportion of scientific software is still developed as standalone scripts and Jupyter notebooks. Such applications often contain implicit execution order, hidden execution state, undeclared software dependencies, and tightly coupled implementations that complicate reproducibility, portability, and reuse \cite{pimentel2019}.

Several studies have investigated techniques for migrating notebooks to cloud environments. Cunha \textit{et al.} proposed context-aware execution migration to preserve notebook state during migration, while demonstrating that block-cell migration outperforms single-cell migration for computationally intensive notebooks \cite{cunha2021}. Duan and Dennis introduced Jup2Kub, which translates Jupyter notebook
pipelines into fault-tolerant distributed deployments on Kubernetes
\cite{duan2023}. NaaVRE further supports notebook migration by automatically extracting workflow components from notebook cells and containerising them for execution within collaborative cloud environments \cite{zhao2022naavre}.

These approaches considerably improve notebook portability and execution. However, they primarily assume that suitable workflow boundaries are already known or are manually defined. Determining how legacy scientific software should be decomposed into reusable workflow components remains largely dependent on expert knowledge.

\subsection{Architectural Analysis for Workflow Generation}

Software architecture recovery has long been used to understand and modernise legacy software systems before restructuring or migration. RM-ODP provides a standard architectural framework that analyses distributed systems from complementary enterprise, information, computational, engineering, and technology viewpoints \cite{ISO10746}. Considering these viewpoints enables software systems to be understood from multiple architectural perspectives rather than solely through their source-code structure.

Dependency analysis provides another complementary perspective for software modularisation. DSM represents dependencies between software elements and has been widely applied to identify cohesive modules, evaluate coupling, and support software restructuring \cite{browning2001,eppinger2012}. Because DSM provides quantitative measures of dependency structure, it offers an objective basis for assessing alternative software decompositions.

Recent advances in LLMs have demonstrated remarkable capabilities for program understanding, code summarisation, documentation generation, and software transformation. These capabilities make LLMs promising tools for analysing heterogeneous legacy software artefacts and assisting workflow generation.

Despite these advances, workflow migration, software architecture recovery, and AI-assisted software engineering have largely evolved as separate research directions. Existing notebook migration approaches primarily focus on execution and deployment, architectural methods recover software structure without generating executable workflows, and LLM-based software engineering generally lacks explicit architectural guidance and objective structural quality assessment.

Consequently, there remains a gap between understanding legacy scientific software and automatically transforming it into reusable scientific workflows. This paper addresses this gap by integrating RM-ODP-guided architectural analysis, LLM-based workflow decomposition, and DSM-based structural quality assessment into a single iterative framework for generating reusable workflows from legacy scientific applications.

\section{Research Methodology}

\begin{figure*}[t]
    \centering
    \includegraphics[width=0.8\textwidth]{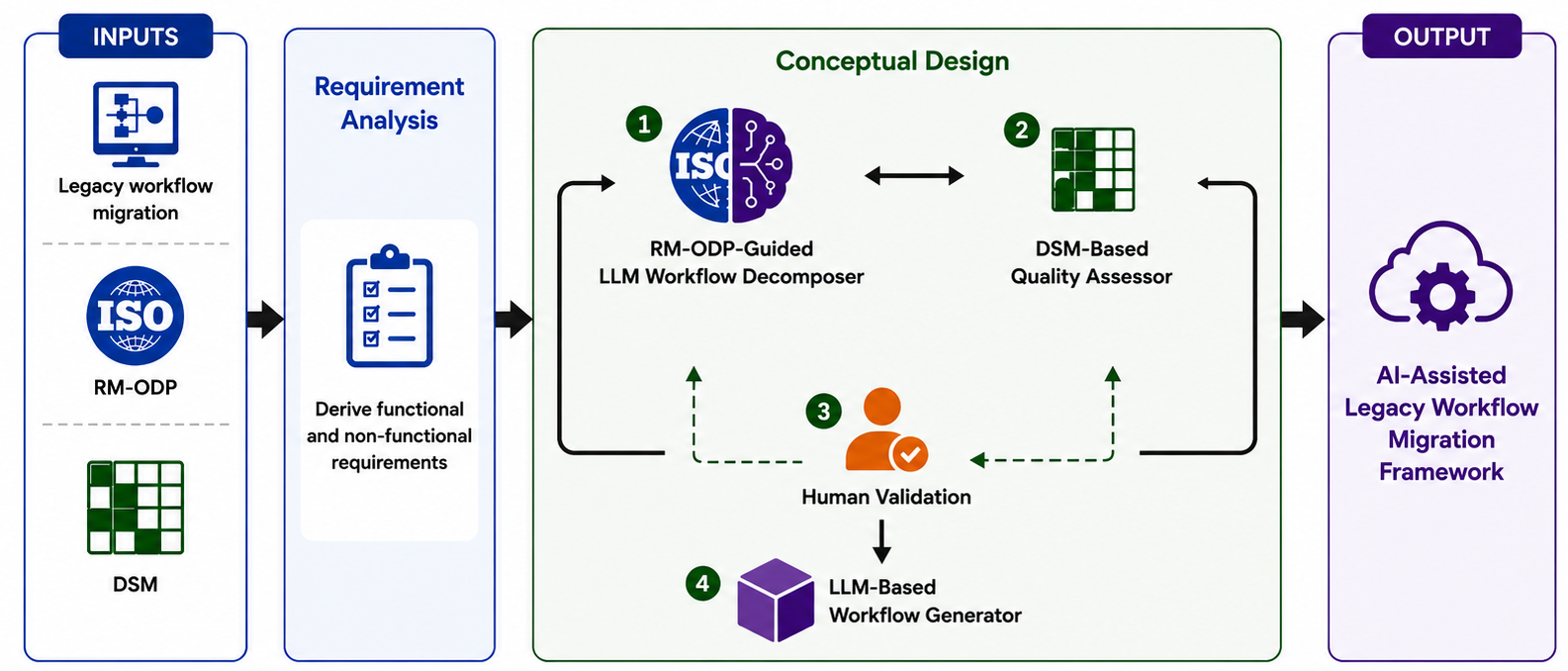}
    \caption{Overview of the two-stage research methodology used to derive the proposed AI-assisted Legacy Workflow Migration Framework}
    \label{fig:methodology}
\end{figure*}

This study adopts a Design Science Research (DSR) methodology to develop an AI-assisted framework for migrating legacy scientific applications into reusable cloud-native workflows. Rather than proposing the framework directly, the architecture was systematically derived by identifying the requirements of legacy workflow migration and selecting appropriate software engineering methods to satisfy those requirements. The overall research methodology is illustrated in Fig.~\ref{fig:methodology}.

The methodology comprises two stages. First, it analyses the challenges of migrating legacy scientific applications to derive the functional and non-functional requirements for the framework. Second, it translates these requirements into a conceptual migration framework by integrating complementary software engineering methods and AI techniques. The resulting design is presented as the proposed AI-assisted Legacy Workflow Migration Framework in the following section.

\subsection{Requirement Analysis}

The first stage derives the design requirements for the proposed AI-assisted legacy workflow migration framework by analysing recurring challenges in legacy workflow migration, scientific workflow engineering, and cloud-based VREs. Existing approaches are limited by incomplete architectural documentation, implicit software dependencies, tightly coupled implementations, limited component modularity, poor reproducibility, and largely manual migration processes that hinder the decomposition of legacy scientific applications into reusable workflow components. Based on these challenges, functional and non-functional requirements were derived to guide the framework design. The functional requirements define the capabilities needed for AI-assisted workflow migration, while the non-functional requirements specify the desired quality attributes of the resulting reusable workflows. Table~\ref{tab:requirements} summarizes the identified requirements.

\begin{table}[t]
\centering
\caption{Functional and non-functional requirements derived for legacy workflow migration.}
\label{tab:requirements}
\footnotesize
\begin{tabularx}{\columnwidth}{p{0.12\columnwidth}X}
\hline
\textbf{ID} & \textbf{Requirement} \\
\hline
\multicolumn{2}{l}{\textbf{Functional Requirements}} \\
\hline
FR1 & Analyse legacy applications to extract architectural information, computational tasks, execution flow, and software dependencies. \\
FR2 & Automatically decompose legacy applications into reusable workflow components. \\
FR3 & Preserve the scientific functionality and execution semantics of the original application throughout migration. \\
FR4 & Generate executable workflow components and workflow specifications suitable for containerization and deployment in cloud-based VREs. \\
FR5 & Support iterative workflow refinement based on structural quality assessment and expert feedback. \\
\hline
\multicolumn{2}{l}{\textbf{Non-functional Requirements}} \\
\hline
NFR1 & Minimize manual migration effort through AI-assisted automation. \\
NFR2 & Maximize the reusability of generated workflow components. \\
NFR3 & Produce modular workflow components with low coupling and high cohesion. \\
NFR4 & Ensure reproducible execution by explicitly capturing software dependencies and runtime environments. \\
NFR5 & Support scalable, cloud-native workflow deployment. \\
NFR6 & Provide transparent and traceable migration decisions to support expert validation and iterative refinement. \\
\hline
\end{tabularx}
\end{table}

\subsection{Conceptual Design}

The second stage translates the identified migration requirements into a conceptual framework by integrating complementary software engineering methods that collectively satisfy the functional and non-functional requirements derived in the previous stage. Instead of relying on a single migration technique, the conceptual design integrates architectural analysis, AI-assisted
workflow decomposition, structural quality assessment, expert validation,
and automated workflow generation into a unified migration process.

To analyse and decompose legacy scientific applications (FR1, FR2, and NFR1), the framework incorporates an \textit{RM-ODP-Guided LLM Workflow Decomposer}. Because legacy applications are often poorly documented and lack explicit architectural descriptions, direct workflow extraction from source code alone is error-prone. The decomposer integrates the five RM-ODP viewpoints (Enterprise, Information, Computational, Engineering, and Technology) into the LLM reasoning process, enabling the identification of scientific objectives, information flow, computational responsibilities, deployment characteristics, and technology dependencies before deriving workflow components, interfaces, and execution dependencies. This architecture-guided decomposition improves consistency while reducing manual migration effort.

To improve structural quality (FR5, NFR2, and NFR3), the framework employs a \textit{DSM-Based Quality Assessor}. Although LLMs can generate meaningful workflow decompositions, they do not guarantee desirable software engineering properties such as low coupling, high cohesion, appropriate component granularity, or minimal dependencies. The quality assessor evaluates these structural properties using DSM analysis and produces a quality report that supports iterative refinement of the workflow decomposition.

To preserve scientific correctness and ensure transparent migration decisions (FR3 and NFR6), the framework incorporates \textit{Human Validation}. Domain experts review the workflow after decomposition to verify that it preserves the functionality and execution semantics of the original application, and again after DSM assessment to evaluate both the workflow and the generated quality report. When revisions are required, expert feedback and the DSM report are returned to the RM-ODP-Guided LLM Workflow Decomposer for iterative refinement.

Finally, the framework addresses workflow implementation (FR4, NFR4, and NFR5) through an \textit{LLM-Based Workflow Generator}. Separating workflow design from implementation, the generator receives the validated workflow specification, including the approved workflow schema, mapped legacy modules, component descriptions, interfaces, software dependencies, runtime libraries, and original artifacts. It then generates executable workflow components together with the workflow definition. By explicitly incorporating dependency and environment information, the generated workflow supports reproducible execution while producing containerizable components suitable for deployment within cloud-based VREs.

Together, these design choices satisfy the identified functional and
non-functional requirements and provide the conceptual basis for the
migration framework described in the following section.

\section{Proposed AI-Assisted Legacy Workflow Migration Framework}

\begin{figure}[t]
    \centering
    \includegraphics[width=\columnwidth]{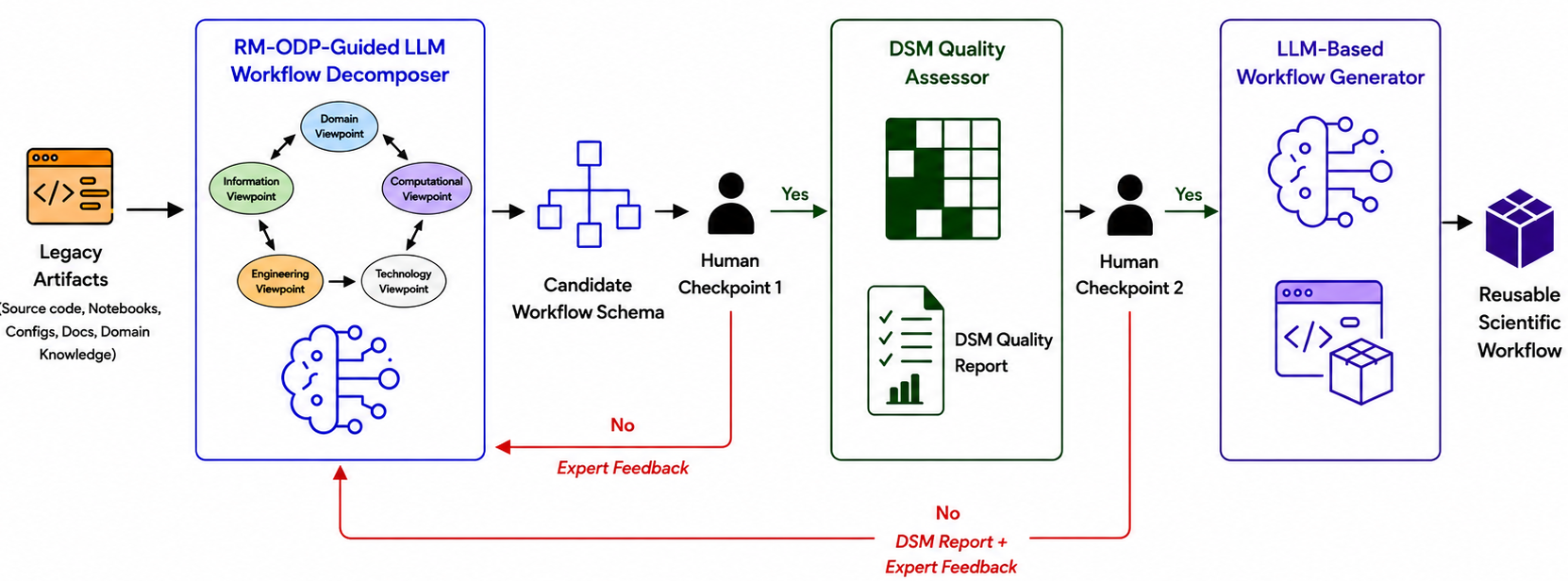}
    \caption{Architecture of the proposed AI-assisted legacy workflow migration framework.}
    \label{fig:architecture}
\end{figure}

Figure~\ref{fig:architecture} illustrates the proposed AI-assisted Legacy Workflow Migration Framework. It comprises three software components: the \textit{RM-ODP-Guided LLM Workflow Decomposer}, the \textit{DSM Quality Assessor}, and the \textit{LLM-Based Workflow Generator}, connected through two expert validation checkpoints that support iterative refinement. The functionality of each component is described in the following subsections.


\subsection{RM-ODP-Guided LLM Workflow Decomposer}

The migration process begins with the \textit{RM-ODP-Guided LLM Workflow Decomposer}, which derives a reusable workflow schema from heterogeneous legacy artifacts using RM-ODP-guided LLM reasoning. Rather than introducing RM-ODP as a separate software component, the five RM-ODP viewpoints are embedded directly into the LLM prompt to support architectural reasoning during workflow decomposition.

The decomposer receives legacy artifacts, including source code, Jupyter notebooks, configuration files, and documentation, together with descriptions of the five RM-ODP viewpoints and the expected workflow-schema format. The prompt guides the LLM to analyse the application from complementary architectural perspectives, including its scientific purpose and stakeholder objectives (Enterprise viewpoint), information flow (Information viewpoint), functional responsibilities and execution dependencies (Computational viewpoint), deployment and orchestration concerns (Engineering viewpoint), and software technologies and runtime environments (Technology viewpoint).

Considering these viewpoints enables the LLM to identify workflow components according to their scientific responsibilities rather than the implementation structure of the source code. The output is a \textit{candidate workflow schema} describing each workflow component, including its identifier, responsibility, mapped legacy modules, inputs, outputs, execution dependencies, required software libraries, container suitability, and implementation metadata for subsequent workflow generation.

Before DSM-based quality assessment, the candidate workflow schema undergoes expert validation to verify that it correctly represents the scientific behaviour of the original application. Domain experts review the identified workflow components, information flows, execution dependencies, deployment considerations, and component boundaries for completeness, scientific correctness, and traceability to the legacy artifacts. If revisions are required, the workflow schema is regenerated using expert feedback while preserving previously validated design decisions; otherwise, the validated workflow schema proceeds to the DSM Quality Assessor.

\subsection{DSM Quality Assessor}

Once the candidate workflow schema has passed the initial scientific validation, it is analysed by the \textit{DSM Quality Assessor}, which evaluates the structural quality of the proposed workflow decomposition using DSM analysis.

The quality assessor analyses dependency relationships among workflow components and evaluates software engineering properties including coupling, cohesion, cyclic dependencies, decomposition granularity, and overall dependency structure. It then produces a structured \textit{DSM Quality Report} describing architectural weaknesses together with recommendations for improving the workflow decomposition. Unlike the workflow decomposer, it evaluates the workflow schema without modifying either the schema or its implementation.

Following the assessment, domain experts review both the workflow schema and the DSM Quality Report to determine whether the decomposition satisfies the required scientific and software engineering quality criteria. If refinement is required, the report and expert feedback are returned to the RM-ODP-Guided LLM Workflow Decomposer, which revises the workflow schema while preserving previously validated scientific functionality. This process is repeated until the workflow schema is approved from both scientific and architectural perspectives.

The approved workflow schema, together with the validated component descriptions, interfaces, execution dependencies, software libraries, and implementation metadata, is then passed to the \textit{LLM-Based Workflow Generator}.

\subsection{LLM-Based Workflow Generator}

The \textit{LLM-Based Workflow Generator} transforms the validated workflow schema into an executable reusable scientific workflow.

The generator receives the approved workflow schema together with the mapped legacy source modules, component descriptions, declared inputs and outputs, execution dependencies, required software libraries, and the original legacy artifacts. Using these inputs, the LLM generates the implementation of each workflow component while preserving the validated architecture. It also produces the corresponding workflow definition specifying the execution order and dependency relationships among the generated components.

By separating workflow decomposition from implementation, the framework ensures that scientific validation and structural quality assessment are completed before code generation, preventing architectural errors from propagating into the executable workflow. The resulting workflow consists of executable components and the corresponding workflow definition, ready for deployment within cloud-based VREs.

\section{Results}

The proposed migration approach was evaluated using the two representative
legacy scientific applications summarized in
Table~\ref{tab:case_studies}. The selected case studies differ in both
their scientific objectives and software structures, providing
complementary scenarios for evaluating the proposed migration approach.

All LLM-assisted stages were implemented using GPT-4o
(\texttt{gpt-4o-2024-11-20}) through the OpenAI API using the default
generation settings. The same model configuration was used for both case
studies and across all refinement iterations. The decomposition prompts
included the legacy artifacts, the five RM-ODP viewpoint descriptions,
and the required workflow-schema format. During refinement, the previous
workflow schema, DSM Quality Report, and expert feedback were additionally
provided to the LLM.

The evaluation considered four aspects. First, the iterative refinement
process was assessed by analysing how feedback improved the generated
workflow decomposition. Second, the final LLM-generated workflows were
compared with manually constructed reference workflows. These reference
workflows were used only for evaluation and were not provided to the LLM
or used to guide the refinement process. Third, the functional correctness
of the migrated workflows was validated by comparing their outputs with
those of the original applications. Finally, workflow component
reusability was evaluated using the reusability metrics proposed by Singh
and Tomar~\cite{singh2014}.

\begin{table}[t]
\centering
\caption{Characteristics of the evaluated legacy scientific applications.}
\label{tab:case_studies}
\footnotesize
\begin{tabularx}{\columnwidth}{p{0.2\columnwidth}p{0.3\columnwidth}p{0.15\columnwidth}p{0.15\columnwidth}}
\hline
\textbf{Case} & \textbf{Domain} & \textbf{Artifact} & \textbf{Source LOC} \\
\hline
BGIS \cite{BGIS} &
Urban digital twin / blue-green infrastructure siting &
Python scripts &
483 \\
\hline
FreVA (Food2Chemical) \cite{Food2Chemical} &
Food-loss-to-chemical resource optimisation &
Jupyter notebook &
50 \\
\hline
\end{tabularx}
\vspace{1mm}
\raggedright
\footnotesize
\textit{Note:} Source LOC denotes the number of lines of source code in
the original legacy application.

\end{table}

\subsection{Workflow Refinement Across Iterations}

The proposed framework supports iterative workflow refinement through
feedback generated at Decision Point~2. After each workflow generation
step, the DSM-based quality assessment evaluates the candidate workflow
and produces recommendations when the workflow does not satisfy the
quality criteria. These recommendations are incorporated into the next
generation iteration until an acceptable workflow is obtained or the
maximum number of iterations is reached.

Table~\ref{tab:workflow_refinement} summarizes the refinement process for
the two case studies.
\begin{table*}[t]
\centering
\caption{Workflow refinement across iterations .}\label{tab:workflow_refinement}
\footnotesize
\begin{tabular}{p{0.05\textwidth}c c c p{0.35\textwidth}p{0.2\textwidth}c}
\hline
\textbf{Case} &
\textbf{Iter.} &
\textbf{GT \#} &
\textbf{Gen. \#} &
\textbf{Issues Identified} &
\textbf{Refinement Performed} &
\textbf{Status} \\
\hline

BGIS &
1 &
5 &
6 &
One additional workflow component and an incorrect dependency between two components detected by DSM analysis. &
Merged redundant component and corrected dependency relationships. &
Rejected \\

&
2 &
5 &
5 &
No critical semantic or structural inconsistencies remained. &
-- &
Accepted \\

\hline

FreVA &
1 &
3 &
5 &
Several scientific responsibilities assigned to incorrect workflow components. &
Reassigned component responsibilities based on semantic feedback. &
Rejected \\

&
2 &
3 &
2 &
Semantic decomposition accepted, but DSM detected merged solver responsibilities and suboptimal component boundaries. &
Separated solver responsibilities and refined component boundaries. &
Rejected \\

&
3 &
3 &
3 &
No semantic or structural inconsistencies detected. &
-- &
Accepted \\

\hline
\end{tabular}
\raggedright
\footnotesize \\ Note: GT \# = number of reference workflow components; Gen. \# = number of generated workflow components.
\end{table*}

\subsection{Generated Reusable Workflows}

The proposed migration methodology successfully transformed both legacy
applications into reusable containerized workflows executable in the
NaaVRE Virtual Lab. NaaVRE was selected as the target platform because
it supports cloud-native workflow orchestration using reusable workflow
components. Despite differences in implementation and scientific domain,
both applications were decomposed into modular components with explicit
interfaces and execution dependencies.

Table~\ref{tab:generated_workflows} compares the manually constructed
reference workflows with the final LLM-generated workflows. BGIS,
originally implemented as two standalone Python scripts, was decomposed
into five workflow components corresponding to the major stages of the
blue-green infrastructure analysis pipeline. FreVA, implemented as a
Jupyter notebook, was decomposed into three workflow components
representing data loading, optimization, and result generation.

\begin{table*}[t]
\centering
\caption{Comparison between the manually constructed (ground-truth) workflows and the final LLM-generated workflows.}
\label{tab:generated_workflows}
\footnotesize
\begin{tabular}{p{0.10\textwidth}p{0.36\textwidth}p{0.36\textwidth}c}
\hline
\textbf{Case} &
\textbf{Ground-Truth Components} &
\textbf{LLM-Generated Components} &
\textbf{Equivalent} \\
\hline
BGIS &
Data Loader,
Land Cover Mapper,
Feasibility Checker,
Output Generator,
Optimizer &
DataParser,
LandUseLoader,
FeasibilityChecker,
DataIntegrator,
OptimizationSolver &
Yes \\
\hline
FreVA &
Data Loading,
Optimization Model,
Result Generation &
DataProcessing,
ModelBuilder,
ResultDisplay &
Yes \\
\hline
\end{tabular}
\end{table*}

The generated workflows closely matched the reference decompositions for both case studies. In FreVA, the optimization model construction and solver execution were combined into a single workflow component because the internal Pyomo model cannot be reliably transferred between containers. This design preserves correct execution while maintaining workflow reusability.

Each generated workflow exposes configurable inputs and parameters
through explicit component interfaces rather than relying on hard-coded
values. This allows the same workflow to be reused with different
datasets, optimization objectives, and execution settings without
modifying the underlying source code. Consequently, both applications
were transformed from standalone research software into reusable
scientific workflows that can be shared, executed, and reproduced
through the Virtual Lab.

\subsection{Workflow Validation}

To evaluate whether the migration preserved the scientific functionality of the original applications, the outputs generated by the reusable workflows were compared with those produced by the legacy implementations.

For BGIS, the migrated workflow was executed on both a small validation dataset and a real-world dataset containing approximately 72,000 land parcels. The generated GeoPackage files, feasibility classifications, optimization results, and visualizations were identical to those of the original Python implementation. For FreVA, the migrated workflow reproduced the same optimization objective values and generated identical reports for both supported optimization objectives. In both case studies, no differences were observed between the outputs of the legacy applications and the migrated workflows, confirming that the proposed approach preserved the scientific functionality and execution semantics of the original applications despite changes to the execution environment and workflow architecture.

\subsection{Component Reusability Evaluation}
\label{sec:component_reuse}

In addition to comparing the generated workflows with the reference
decompositions, we evaluated the reusability of the generated workflow
components using the model proposed by Singh and
Tomar~\cite{singh2014}. The evaluation considers four quality
attributes: Interface Complexity (IC), Understandability (UC), Rate of
Component Customisability (RCC), and Reliability (R).

Because the original model was developed for black-box software
components, the metrics were adapted to workflow components. IC measures
the complexity of a component interface based on its inputs, outputs,
and external dependencies, where lower values indicate simpler
interfaces. UC evaluates the completeness and clarity of the component
specification. RCC measures the proportion of configurable execution
parameters relative to hard-coded values, reflecting the ease with which
a component can be adapted to different workflows. The final metric
measures the proportion of successful executions during workflow
validation.

The four metrics were computed for every component in the reference and
LLM-generated workflows and averaged to obtain workflow-level scores.
Table~\ref{tab:reusability_metrics} summarizes the results.

\begin{table}[t]
\centering
\caption{Average component reusability metrics for the reference and LLM-generated workflows.}
\label{tab:reusability_metrics}
\footnotesize
\begin{tabularx}{\columnwidth}{lXcccc}
\hline
\textbf{Case} & \textbf{Workflow} & \textbf{IC$\downarrow$} & \textbf{UC$\uparrow$} & \textbf{RCC$\uparrow$} & \textbf{R$\uparrow$} \\
\hline
BGIS  & Reference & 0.23 & 0.91 & 0.82 & 1.00 \\
       & LLM       & 0.25 & 0.89 & 0.80 & 1.00 \\
\hline
FreVA & Reference & 0.19 & 0.94 & 0.86 & 1.00 \\
       & LLM       & 0.20 & 0.92 & 0.84 & 1.00 \\
\hline
\end{tabularx}
\end{table}

\section{Discussion}

The evaluation demonstrates that the proposed RM-ODP-guided workflow
migration approach can successfully transform legacy scientific
applications with different implementation styles into reusable
cloud-native workflows. Despite differences in programming paradigm,
application domain, and software structure, both BGIS and FreVA were
decomposed into workflow components that reproduced the behaviour of the
original applications while exposing explicit interfaces suitable for
workflow composition.

The workflow refinement results further indicate that combining LLM-based
generation with iterative DSM-based quality assessment improves the
quality of the generated workflow structures. Initial workflow
decompositions contained structural issues such as inappropriate
component boundaries and incorrect dependencies. Incorporating feedback
from the DSM assessment and expert validation enabled the LLM to
progressively refine the decomposition until these issues were resolved.
The final workflows also showed close agreement with the independently
constructed reference workflows, which were used only for evaluation.
This suggests that architectural quality assessment provides an effective
feedback mechanism for guiding LLM-based workflow generation rather than
relying on a single-generation step.

The evaluation also highlights that successful migration depends on more
than identifying computational modules. Across both case studies,
migration required resolving implicit assumptions embedded in legacy
software, including undeclared software dependencies, hard-coded
filesystem paths, machine-specific configuration, and hidden execution
dependencies. These issues are rarely represented explicitly in source
code but substantially affect reproducibility once applications are
executed on shared cloud infrastructures. Consequently, migration should
be viewed as both a software decomposition problem and an environment
reconstruction problem.

\subsection{Lessons Learned}

Several observations consistently emerged from the migration of both
applications.

First, reproducible execution requires explicit specification of the
software environment. Legacy scientific applications frequently assume
that libraries and external tools are already available on the
developer's machine. Containerization therefore becomes not only a
deployment mechanism but also a way to preserve the complete execution
environment required for reproducibility.

Second, legacy applications commonly embed assumptions about the local
filesystem through hard-coded paths and machine-specific configuration.
Replacing these assumptions with parameterized workflow inputs and
persistent cloud storage substantially improves workflow portability and
allows the same workflow to execute across different research
infrastructures without modifying the source code.

Third, meaningful workflow decomposition should reflect scientific
responsibilities rather than source-code structure alone. In BGIS, the
identified workflow components correspond to stages of the urban
planning analysis, while FreVA was decomposed according to the logical
optimization process instead of notebook cell boundaries. This supports
the use of RM-ODP viewpoints for identifying reusable workflow
components from complementary architectural perspectives rather than
purely syntactic code analysis.

Finally, workflow decomposition must balance modularity with practical
implementation constraints. Although finer-grained components generally
improve reuse, tightly coupled computations cannot always be separated
without introducing unnecessary complexity or excessive communication
overhead. For example, the optimization model construction and solver
execution in FreVA remained within a single workflow component because
the internal Pyomo model could not be reliably serialized across
container boundaries. This illustrates that architectural decomposition
should consider both scientific modularity and execution constraints.

\subsection{Limitations}

This work has several limitations. First, the evaluation includes two
representative legacy scientific applications. Although they originate
from different scientific domains and represent different legacy
artifacts (Python scripts and Jupyter notebooks), additional case
studies are needed to evaluate the generality of the proposed approach.
Second, the reference workflows used for evaluating generation accuracy
were manually constructed, introducing some degree of expert judgement.
Third, the proposed framework currently relies on a human expert to
review workflow decompositions during iterative refinement. Reducing this
dependency through more automated architectural validation represents an
important direction for future work. Finally, the evaluation focused on
functional correctness, workflow quality, and execution performance;
future studies should also investigate long-term maintainability and
reuse by independent researchers.

\section{Conclusion and Future Work}

This paper presented an AI-assisted framework for migrating legacy
scientific applications into reusable cloud-native workflows. The
framework combines RM-ODP-guided architectural analysis, LLM-based
workflow decomposition, and DSM-based quality assessment to automate
workflow migration while iteratively improving workflow quality.

The evaluation on two legacy scientific applications demonstrated that
the migrated workflows preserved the functionality of the original
applications while producing reusable workflow components suitable for
cloud execution. Iterative DSM-guided refinement consistently improved
agreement with reference workflows and enhanced workflow modularity,
highlighting the value of combining architectural guidance with
LLM-based generation.

Future work will evaluate the framework on a broader range of legacy
applications and programming languages to assess its generality.
Additional directions include reducing reliance on expert feedback
through automated architectural validation and self-refinement, and
supporting richer workflow constructs, such as loops and conditional
execution, together with interoperability across multiple scientific
workflow platforms.
\section*{Acknowledgment}
This research was partially supported by the Dutch Research Council (NWO) Large-Scale Research Infrastructures (LSRI) programme for the LTER‑LIFE infrastructure (grant 184.036.014), LifeWatch ERIC, and the European Union through the projects ENVRI‑Hub Next (101131141), EVERSE (101129744), OSCARS (101129751), and BMD (101181294).

\end{document}